\documentclass[b5paper,11pt,twoside]{article}

\usepackage[b5paper,top=26mm,bottom=25mm]{geometry}
\usepackage[utf8x]{inputenc}
\usepackage{default}
\usepackage{amsmath}
\usepackage{amsfonts}
\usepackage{amscd}
\usepackage{amssymb}
\usepackage{amsthm}
\usepackage{mathrsfs}
\usepackage{stmaryrd}
\usepackage{t1enc}
\usepackage[polish]{babel}
\usepackage{ucs}
\usepackage{makeidx}
\usepackage{times}
\usepackage{graphics}
\usepackage{listings}
\title{\Large \textbf{GENERATOR KODÓW LINIOWYCH O SKOŃCZONYCH CHARAKTERYSTYKACH}}

\author{{\normalsize Mariusz Frydrych, Wojciech Horzelski}\\
  {\small Uniwersytet Łódzki, Wydział Matematyki i Informatyki}\\
  {\small frydrych@math.uni.lodz.pl, horzel@math.uni.lodz.pl}\\
 }

\date{}
\newtheorem*{df}{Definicja}

\newtheorem*{stw}{Stwierdzenie}

\newtheorem*{wn}{Wniosek}

\begin{document}

\sloppy
\maketitle
\pagenumbering{arabic}
\pagestyle{empty}
\thispagestyle{empty}

\noindent{\small \textbf{Streszczenie} 
Praca opsuje metodę szybkiego generowania kodów liniowych w wymiarze ,,połówkowym'', tzn. gdy wymiar kodu jest równy jego kowymiarowi. 
Kod jest reprezentowany w przestrzeni wektorowej nad ciałem skończonym charakterystyki większej niż dwa, co dało mozliwość wykorzystania 
automorfizmu Frobeniusa do konstrukcji pewnych operatorów liniowych mających naturę geometryczną. 
Metodę zilustrowano przykładem w wymiarze trzy (wymiar i kowymiar kodu) nad ciałem charakterystyki siedem. 

}

\noindent{\small \textbf{Słowa kluczowe}: kody liniowe, kodowanie, Grassmanian}


\section{Wprowadzenie}
Kody liniowe stosowane są powszechnie w przesyłaniu danych w zaszumionym medium transmisyjnym. 
Przez wymiar kodu rozumie się przepustowość łącza dla transitowanej informacji,
z kolei kowymiar kodu, mierzy tak zwaną nadmiarowość czyli ilo informacji niezbędnej do wykrywania i ewentualnej korekcji błędów w przesyłanych danych.
Liniowość kodu zankomicie upraszcza procesy kodowania i dekodowania, co skutkuje dużą wydajnością implementowanych algorytmów.
Z oczywistych powodów, najczęściej stosuje się kody binarne, co znacznie zawęża spektrum możliwej do uzyskania jakości kodu.
Zastosowanie większej liczby stanów (ciał skończonych charakterystyki większej niż dwa), daje elastyczną strukturę kolekcji kodów liniowych.

\section{Kod liniowy}

Niech, ~$k, n, p, w, q \in\mathbb{N}, ~~p~ - \textrm{liczba pierwsza}, ~~q=p^w, ~k\leqslant n$.
\newline

\begin{df}
Każdą ~$k$-wymiarową podprzestrzeń wektorową ~$C$~ przestrzeni ~$n$-wymiarowej ~$\mathbb{F}_q^n$~ nazywamy 
\emph{kodem liniowym} o długości $n$, wymiaru ~$k$, ~nad ciałem ~$\mathbb{F}_q$ ~~(\cite{SB}, \cite{VP}).
\end{df}

Wybór bazy ~~$B=(b_1,\ldots, b_k), ~b_1,\ldots,b_k\in C \subset \mathbb{F}_q^n$~ indukuje monomorfizm przestrzeni liniowych
\begin{displaymath}
\iota \colon \mathbb{F}_q^k  \longrightarrow \mathbb{F}_q^n, ~~ \left( \begin{array}{c}
               \xi_1\\
		\vdots\\
		\xi_k
              \end{array}
\right) \mapsto \sum_{j=1}^{k} \xi_j b_j, ~~\operatorname{im}(\iota) = C.
\end{displaymath}

zwany \emph{kodowaniem liniowym}.


Dostajemy 
\begin{displaymath}
0 \longrightarrow \mathbb{F}_q^k \xrightarrow{~~\iota~~} \mathbb{F}_q^n \xrightarrow{~~\pi~~} \mathbb{F}_q^n/C \longrightarrow 0
\end{displaymath}

tzw. krótki ciąg dokładny przestrzeni wektorowych. ~$\operatorname{codim} C = \dim \mathbb{F}_q^n/C = n-k$. ~
Składając ~$\pi$~ z dowolnym izomorfizmem ~~$\mathbb{F}_q^n/C \xrightarrow{~~\approx~~} \mathbb{F}_q^{n-k}$~~
dostajemy ponownie (krótki ciąg dokładny). Operator (macierz) ~$H$~ nazywamy \emph{anihilatorem, macierzą kontrolną (check matrix)} kodu ~$C$.

\begin{displaymath}
 0 \longrightarrow \mathbb{F}_q^k \xrightarrow{~~\iota~~} \mathbb{F}_q^n \xrightarrow{~~H~~} \mathbb{F}_q^{n-k} \longrightarrow 0
\end{displaymath}


  Kowymiar podprzestrzeni ~$\operatorname{codim} C = n-k$~ to ,,ilość stopni kontrolnych kodu ~-~ nadmiarowość'' ~ 
  a wymiar ~$\dim C = k$~ ,,zawartość informacji''.
\newline

Wektory bazowe $b_1,\ldots,b_k \in \mathbb{F}_q^n$ są liniowo niezależne, więc znajdziemy podciąg  
$1\leqslant j_1 < \ldots < j_k  \leqslant n$, taki że macierz

\begin{displaymath}
b_{j_1,\ldots,j_k} = \left[
 \begin{matrix}
  b_{j_1,1} & \ldots &  b_{j_1,k}\\
  \vdots & \vdots &  \vdots \\
  b_{j_k,1} & \ldots &  b_{j_k,k} 
 \end{matrix}
\right]
\end{displaymath}

jest nieosobliwa.


\begin{displaymath}
 P \cdot B \cdot b_{j_1,\ldots,k_k}^{-1} = 
\left[
 \begin{matrix}
  1 & \ldots & 0\\
  \vdots & \ddots & \vdots \\
  0 & \ldots & 1 \\
  a_{1,1} & \ldots & a_{1,k} \\
  \vdots & \ddots & \vdots \\
  a_{n-k,1} & \ldots & a_{n-k,k} 
 \end{matrix}
\right] 
\end{displaymath}

$P$~~ jest odpowiednią macierzą permutacji osi współrzędnych przestrzeni ~~$\mathbb{F}_q^n$.
~Jest to tzw. \emph{standardowa} postać bazowa kodu liniowego ~$C$.


Dla postaci standardowej kodu ~$C$~
 \begin{displaymath}
B = \left[\begin{array}{c}
I_{k,k}\\
A
\end{array}\right]
\end{displaymath}

anihilator (macierz kontrolna, check matrix) ~~$H$~~ ma postać

\begin{displaymath}
H = \left[\begin{array}{cc}
-A & I_{n-k,n-k}
\end{array}\right].
\end{displaymath}

gdzie ~$I_{k,k}, ~I_{n-k,n-k}$~ są macierzami jednostkowymi odpowiednich wymiarów.

\subsection{Grassmanian}

\begin{df}

Ogół wszystkich podprzestrzeni ~$k$-wymiarowych przestrzeni ~$n$-wymiarowej ~$\mathbb{F}_q^n$~ nazywamy 
\emph{rozmaitością Grassmana} lub \emph{Grassmanianem} i oznaczamy 

\begin{displaymath}
 Grass( k, n, \mathbb{F}_q ) = \{ V\colon V\subset \mathbb{F}_q^n \wedge \dim_{\mathbb{F}_q} V = k \}.
\end{displaymath}

\end{df}

 Z postaci standardowej widać, że Grassmanian ~$Grass( k, n, \mathbb{F}_q )$~ jest rozmaitością wymiaru ~$k\cdot (n-k)$ 
nad ciałem ~$\mathbb{F}_q$~ i można go naturalnie zanurzyć jako kwadrykę w przetrzeni rzutowej 

\begin{displaymath}
 Grass( k, n, \mathbb{F}_q ) \hookrightarrow \mathbb{P}( \varLambda^k \mathbb{F}_q^n )
\end{displaymath}

\begin{displaymath}
\operatorname{span}( v_1, \ldots, v_k ) \mapsto\operatorname{span}( v_1\wedge \ldots \wedge v_k ).
\end{displaymath}


Pełna grupa liniowa ~$GL(\mathbb{F}_q^n)$~ działa tranzytywnie na podprzestrzeniach ustalnego wymiaru, stąd 

\begin{stw}
 Grassmanian jest przestrzenią jednorodną 

\begin{displaymath}
 Grass( k, n, \mathbb{F}_q ) \simeq GL(\mathbb{F}_q^n) / F( k, n, \mathbb{F}_q )
\end{displaymath}

 ~$F( k, n, \mathbb{F}_q )$~ jest grupą macierzy postaci

\begin{displaymath}
 \left[
 \begin{array}{cc}
  a &  b\\
  0 &  c
 \end{array}
\right]
\end{displaymath}

gdzie 

\begin{displaymath}
 a \in GL(\mathbb{F}_q^k ), ~~~c \in GL(\mathbb{F}_q^{n-k}), ~~~ b \in M( k, n-k, \mathbb{F}_q ). 
\end{displaymath}

~$b$~ jest dowolną macierzą prostkątną o ~$k$ ~wierszach i ~$n-k$ ~kolumnach i elemntach w ciele ~$\mathbb{F}_q$.
\end{stw}


Ponieważ grupa liniowa składa się z 
\begin{align*}
 \# ~~GL( \mathbb{F}_q^n ) &= (q^n-1)(q^n-q)(q^n-q^2)\cdots (q^n-q^{n-1}) \\
 &=(q^n-1)(q^{n-1}-1)\cdots (q-1)\cdot q^{\binom{n}{2}}
\end{align*}
elementów, otrzymujemy 

\begin{wn}
  Liczba elementów rozmaitości Grassmana wynosi 
\begin{align*}
 \# ~~Grass( k, n, \mathbb{F}_q ) = \dfrac{(q^n-1)(q^{n-1}-1)\cdots (q^{n-k+1}-1)}{(q^k-1)(q^{k-1}-1)\cdots (q-1)}.
\end{align*}
\end{wn}


Poniżej przedstawiono kilka przykładów zestawienia wymiaru kodu( ~$k$),  długości kodu (~$n$), ilości elementów w ciele $\mathbb{F}_q$ (~$q$) z  ilością elementów Grassmanianu (~$\# ~[k,n]_q$).

 \begin{displaymath}
\begin{array}{rrrr}
k & n & q & \# ~[k,n]_q\\
\\

4 & 7 & 2^4 & 301\,490\,686\,407\,185\\
4 & 8 & 2^4 & 19\,758\,795\,115\,067\,683\,345\\
8 & 16 & 2 & 63\,379\,954\,960\,524\, 853\,651\\
3 & 6 & 7^2 & 1\,663\,045\,363\,565\,300
\end{array}
\end{displaymath}

%

\section{Generowanie kodów}

Rozważmy teraz liczbę pierwszą $p>2$, oraz liczby naturalne 
$k,~n=2k$. Będziemy poszukiwać $k$-wymiarowych kodów liniowych
o długości $n$, tzn. długość kodu będzie równa podwojonemu wymiarowi.
Grassmanian ~$Grass(k, 2k, \mathbb{F}_p)$~ jest ,,najbogatszy w wymiarze połówkowym''
bowiem składa się z  
\begin{displaymath}
 \dfrac{(p^{2k}-1)(p^{2k-1}-1)\cdots (p^{k+1}-1)}{(p^k-1)(p^{k-1}-1)\cdots (p-1)}
\end{displaymath}
elementów.

Przestrzeń wekorową $\mathbb{F}_p^n$ nad ciałem $\mathbb{F}_p$ możemy potraktować jako ciało $\mathbb{F}_{p^n}$ 
poprzez rozszerzenie stopnia $n$ ciała prostego $\mathbb{F}_p$ za pomocą nieprzywiedlnego 
wielomianu $f\in\mathbb{F}_p[X], ~\deg f = n$.

Od tej pory będziemy w powyższy sposób utożsamiać ciało $\mathbb{F}_{p^n}$ z 
przestrzenią liniową $\mathbb{F}_p^n$ nad ciałem $\mathbb{F}_p$: 
\begin{displaymath}
 \mathbb{F}_{p^n}  \simeq_f  \mathbb{F}_p^n.
\end{displaymath}

Rozważmy automorfizm Frobeniusa
\begin{displaymath}
\sigma: \mathbb{F}_p^n \to \mathbb{F}_p^n, ~~\sigma(x)= x^p, 
\end{displaymath}
którego $n$-ta iteracja 
\begin{displaymath}
\sigma^n(x)= x^{p^n}
\end{displaymath}
jest identycznością ($Id=1$) na $\mathbb{F}_p^n$.
Ponieważ $n=2k$, to $k$-ta iteracja
$\sigma^k(x)= x^{p^k}$ jest inwolucją. Oznaczmy ją przez $\tau$. 

Otrzymaliśmy operator liniowy
\begin{displaymath}
\tau: \mathbb{F}_p^n \to \mathbb{F}_p^n, ~~\tau^2= 1, 
\end{displaymath}
który w naturalny sposób rozkłada przestrzeń $\mathbb{F}_p^n$ na sumę prostą dwóch podprzestrzeni własnych:
\begin{displaymath}
 V^{+} = \ker (\tau-1), ~~V^{-} = \ker (\tau+1)
\end{displaymath}
\begin{displaymath}
\mathbb{F}_p^n =  V^{+}\oplus  V^{-}
\end{displaymath}

Ponieważ charakterystka ciała jest różna od dwóch, dostajemy dwa operatory idempotentne (rzuty) 
\begin{displaymath}
\pi^{+}, \pi^{-}: \mathbb{F}_p^n \to \mathbb{F}_p^n
\end{displaymath}
\begin{displaymath}
\pi^{+}= \frac{1}{2}(1-\tau), ~~\pi^{-}= \frac{1}{2}(1+\tau), 
\end{displaymath}
spełniające warunki: 
\begin{displaymath}
\pi^{+} + \pi^{-} = 1, 
\end{displaymath}
\begin{displaymath}
\pi^{+} \pi^{-} = 0 = \pi^{-} \pi^{+}, 
\end{displaymath}
\begin{displaymath}
\ker\pi^{+}= \operatorname{im}\pi^{-}= V^{+}, 
\end{displaymath}
\begin{displaymath}
\ker\pi^{-}= \operatorname{im}\pi^{+}= V^{-}.
\end{displaymath}

Z drugiej strony zauważmy, że
\begin{displaymath}
V^{+} = \ker\pi^{+} = \ker (1-\tau) = \ker (1-\sigma^k), 
\end{displaymath}
co oznacza, że $V^{+}$ jest rozszerzeniem stopnia $k$ ciała prostego $\mathbb{F}_p$, tzn. jest izomorficzne
z ciałem skończonym $p^k$-elementowym ~$\mathbb{F}_{p^k}$.

Podsumowując, otrzymaliśmy ciąg kolejnych ciał, rozszerzeń ciała prostego $\mathbb{F}_p$: 
\begin{displaymath}
\mathbb{F}_p \varsubsetneq V^{+} \varsubsetneq \mathbb{F}_p^n
\end{displaymath}
gdzie
\begin{displaymath}
V^{+} \simeq \mathbb{F}_{p^k}, ~~\mathbb{F}_p^n \simeq \mathbb{F}_{p^n}.
\end{displaymath}
\begin{displaymath}
|V^{+} : \mathbb{F}_p|=k, ~|\mathbb{F}_p^n : V^{+}|=2, ~|\mathbb{F}_p^n : \mathbb{F}_p|=n=2k.
\end{displaymath}

Kluczowe dla naszej konstrukcji jest rozszerzenie stopnia dwa $\mathbb{F}_p^n / V^{+}$ ciała $p^k$-elementowego $V^{+}$ przez ciało $p^n$-elementowe $\mathbb{F}_p^n$.
Mianowicie, traktujemy ciało $\mathbb{F}_p^n$ jako dwuwymiarową przestrzeń wektorową nad ciałem $V^{+}$. 
Automofizm ciała $\mathbb{F}_p^n$ jako operator liniowy nad ciałem prostym $\mathbb{F}_p$
\begin{displaymath}
 \tau: \mathbb{F}_p^n \to \mathbb{F}_p^n
\end{displaymath}
jest niezmienniczy na podprzestrzeni  $V^{+}$, więc możemy go traktować jako operator liniowy nad ciałem   $V^{+}\simeq\mathbb{F}_{p^k}$.

Jeżeli wybierzemy dowolny element $\xi\in V^{-}, \xi\neq 0$ to możenie przez $\xi^{-1}$ ustala izomorfizm pomiędzy podprzestrzeniami: 
\begin{displaymath}
 \xi^{-1}: V^{-} \to V^{+}, ~~\xi^{-1}(x)= \xi^{-1}\cdot x
\end{displaymath}
a izomorfizmem odwrotnym jest: 
\begin{displaymath}
 \xi: V^{+} \to V^{-}, ~~\xi(x)= \xi\cdot x
\end{displaymath}

Ponieważ $\xi^2\in V^{+}, \xi\notin V^{+}$ to $\mathbb{F}_p^n \simeq V^{+}[\xi]$, tzn. element $\xi$ realizuje rozszerzenie 
$\mathbb{F}_p^n / V^{+}$ stopnia dwa.

W powyższy sposób dostajemy rozkład ciała $\mathbb{F}_p^n$ na sumę prostą podprzestrzeni liniowych 
\begin{align*}
\mathbb{F}_p^n \simeq&_\xi V^{+}\oplus V^{+}\\
\mathbb{F}_p^n \ni u \mapsto ( \xi^{-1}\cdot \pi&^{+}u, \pi^{-}u ) \in V^{+}\oplus V^{+}
\end{align*}

Każdej jednowymiarowej (nad $V^{+} \simeq \mathbb{F}_{p^k}$) podprzestrzeni wektorowej odpowiada naturalnie $k$-wymiarowa (nad $\mathbb{F}_p$) 
podprzestrzeń liniowa przestrzeni $\mathbb{F}_p^n$. 

\begin{displaymath}
\varTheta: \mathbb{P}^1(\mathbb{F}_{p^k}) \longrightarrow Grass( k, 2k, \mathbb{F}_p ).
\end{displaymath}

Wykorzystując wsółrzędne jednorodne, prostą rzutową $\mathbb{P}^1(\mathbb{F}_{p^k})$ możemy utożsamić z $\mathbb{F}_{p^k} \cup \{\infty\}$
\begin{displaymath}
 (V^{+}\oplus V^{+})/\mathbb{F}_{p^k} \ni [x, y] \longmapsto \left\{\begin{array}{lll}
 \dfrac{x}{y} & dla & y\neq 0 \\
 \infty      & dla & y=0.
\end{array} \right.
\end{displaymath}

Jawna postać włożenia $\varTheta$ wygląda następująco: 

\begin{displaymath}
 \mathbb{F}_{p^k} \simeq V^{+} \ni x \longmapsto \operatorname{span}_{\mathbb{F}_{p^k}} \{ x + \xi \} \subset \mathbb{F}_p^n
\end{displaymath}

\begin{displaymath}
 \infty \longmapsto V^{+} \subset \mathbb{F}_p^n.
\end{displaymath}


\section{Implementacja metody}
Generator kodów opisaną metodą został zaimplementowany w języku C, przy wykorzystaniu biblioteki algebraicznej Computer Algebra System z Uniwersytetu w Bordeaux. 

Wykorzytano tu ciało skończone $\mathbb{F}_{7^6}$ rzędu $7^6 = 117\,649$
przyjmując następujące wartości parametrów: 
\begin{displaymath}
 p=7, ~k=3, ~n=~2k=6, 
\end{displaymath}
wielomian nieprzywiedlny $f \in\mathbb{F}_7[X] $ stopnia $n=6$: 
\begin{displaymath}
 f(X)= X^6 + X^5 + 2X^4 + X^3 + 5X^2 + 3X + 2, 
\end{displaymath}
generator (pierwiastek pierwotny) $g$ ciała $\mathbb{F}_{7^6}$ rzędu $7^6-1 = 117\,648$
\begin{displaymath}
 g(X)= 3X^5 + 4X^4 + 5X^2 + 2X + 2.
\end{displaymath}

Do obliczeń wykorzystano funkcje biblioteki Computer Algebra System (\textit{ffinit(),ffgen(), ffprimroot(), fforder()})  :
\begin{verbatim}
void init_kody(long prec)	
{
  GEN p1;
  p = pol_x(fetch_user_var("p"));
  k = pol_x(fetch_user_var("k"));
  n = pol_x(fetch_user_var("n"));
  f = pol_x(fetch_user_var("f"));
  t = pol_x(fetch_user_var("t"));
  g = pol_x(fetch_user_var("g"));
  p = stoi(7);
  k = stoi(5);
  n = gmulsg(2, k);
  f = ffinit(p, gtos(n), -1);
  t = ffgen(f, -1);
  g = ffprimroot(t, NULL);
  p1 = fforder(g, NULL);
  {
    GEN j;
    for (j = gen_0; gcmp(j, p1) <= 0; j = gaddgs(j, 1))
      pari_printf("%Ps; %Ps\n", j, gpow(g, j, prec));
  }
  return;
}
\end{verbatim}
Realizacją ciała $\mathbb{F}_{7^6}$ jest ucięta algebra wielomianów: 
\begin{displaymath}
 \mathbb{F}_{7^6} \simeq \mathbb{F}_7[X]/(f).
\end{displaymath}

Obliczenia będziemy prowadzić w uporządkowanej bazie sześciowymiarowej przestrzeni wektorowej 
$\mathbb{F}_{7^6} \simeq\mathbb{F}_7^6$ nad ciałem $\mathbb{F}_7$: 
\begin{displaymath}
 ( X^5, X^4, X^3, X^2, X, 1 ).
\end{displaymath}

Macierze automorfizmu Frobeniusa ~$\sigma: \mathbb{F}_7^6 \rightarrow \mathbb{F}_7^6$~ oraz inwolucja $\tau= \sigma^3$: 
\begin{displaymath}
\sigma = \left[\begin{array}{cccccc}
6 & 0 & 2 & 5 & 6 & 0\\
4 & 5 & 6 & 4 & 1 & 0\\
2 & 3 & 5 & 1 & 3 & 0\\
4 & 2 & 1 & 3 & 2 & 0\\
2 & 4 & 3 & 6 & 1 & 0\\
6 & 6 & 4 & 6 & 2 & 1
\end{array}
\right], ~~~
\tau = \left[\begin{array}{cccccc}
3 & 5 & 5 & 0 & 5 & 0\\
6 & 5 & 0 & 3 & 5 & 0\\
4 & 4 & 6 & 2 & 1 & 0\\
1 & 2 & 5 & 1 & 6 & 0\\
1 & 2 & 5 & 2 & 5 & 0\\
6 & 2 & 3 & 6 & 2 & 1
\end{array}
\right]
\end{displaymath}

operatory rzutu (idempotenty) $\pi^{+}$ i $\pi^{-}$: 
\begin{displaymath}
\pi^{+} = \left[\begin{array}{cccccc}
6 & 1 & 1 & 0 & 1 & 0\\
4 & 5 & 0 & 2 & 1 & 0\\
5 & 5 & 1 & 6 & 3 & 0\\
3 & 6 & 1 & 0 & 4 & 0\\
3 & 6 & 1 & 6 & 5 & 0\\
4 & 6 & 2 & 4 & 6 & 0
\end{array}
\right], ~~~
\pi^{-} = \left[\begin{array}{cccccc}
2 & 6 & 6 & 0 & 6 & 0\\
3 & 3 & 0 & 5 & 6 & 0\\
2 & 2 & 0 & 1 & 4 & 0\\
4 & 1 & 6 & 1 & 3 & 0\\
4 & 1 & 6 & 1 & 3 & 0\\
3 & 1 & 5 & 3 & 1 & 1
\end{array}
\right]
\end{displaymath}

bazy podprzestrzeni ~$V^{+}$~ i ~$V^{-}$~(jako odpowiednie kolumny macierzy):
\begin{displaymath}
V^{+} = \left[\begin{array}{ccc}
6 & 1 & 0\\
5 & 0 & 0\\
1 & 0 & 0\\
0 & 1 & 0\\
0 & 1 & 0\\
0 & 0 & 1
\end{array}
\right], ~~~
V^{-} = \left[\begin{array}{ccc}
3 & 1 & 2\\
0 & 4 & 5\\
6 & 4 & 6\\
1 & 0 & 0\\
0 & 1 & 0\\
0 & 0 & 1
\end{array}
\right]
\end{displaymath}

elementy ~$\xi, ~\xi^{-1} \in V^{-}$: 
\begin{displaymath}
 \xi= 2X^5 + 6X^4 + 5X^3 + 5X^2 + 4, ~~~\xi^{-1}= 3X^5 + 6X^3 + X^2.
\end{displaymath}

Dla ujednolicenia oznaczeń, bazą przestrzeni ~$\mathbb{F}_7^6$ jest
\begin{displaymath}
 ( X^5, X^4, X^3, X^2, X, 1 ) = ( e_1, e_2, e_3, e_4, e_5, e_6 )
\end{displaymath}
oraz baza podprzestrzeni ~$V^{+}$
\begin{align*}
 f_1 &= 6e_1 + 5e_2+e_3\\
 f_2 &= e_1 + e_4 + e_5\\
 f_3 &= e_6.
\end{align*}

Mnożenie w ciele ~$\mathbb{F}_7^6$~ można przedstawić jako tensor
\begin{displaymath}
 \mathbb{F}_7^6 \otimes_{\mathbb{F}_7} \mathbb{F}_7^6 \longrightarrow \mathbb{F}_7^6
\end{displaymath}
\begin{displaymath}
 e_j\cdot e_k = \sum_{l=1}^6 c_{j,k}^l e_l, ~~~j, k= 1, \ldots, 6
\end{displaymath}
gdzie współczynniki ~$c_{j,k}^l\in\mathbb{F}_7$~ są stałymi struktury (mnożenia).

Gdy mnożenie przez lewy czynnik ograniczymy do podprzestrzeni ~$V^{+} \simeq \mathbb{F}_{7^3}$~
to otrzymamy częściowy tensor w postaci trzech macierzy 
\begin{displaymath}
f_1\cdot  = \left[\begin{array}{cccccc}
0 & 6 & 6 & 4 & 6 & 6\\
2 & 6 & 5 & 3 & 3 & 5\\
3 & 0 & 4 & 6 & 1 & 1\\
0 & 2 & 6 & 1 & 5 & 0\\
5 & 2 & 4 & 5 & 3 & 0\\
2 & 2 & 6 & 2 & 2 & 0
\end{array}
\right], 
 ~~f_2\cdot = \left[\begin{array}{cccccc}
 1 & 3 & 3 & 6 & 6 & 1\\
 4 & 4 & 6 & 2 & 5 & 0\\
 1 & 3 & 3 & 4 & 0 & 0\\
 6 & 4 & 6 & 2 & 3 & 1\\
 6 & 0 & 5 & 1 & 4 & 1\\
 1 & 1 & 2 & 2 & 5 & 0
\end{array}
\right], 
 ~~f_3\cdot = Id.
\end{displaymath}

Włożenie generujące ~$7^3+1=344$~ kodów liniowych  
\begin{displaymath}
\varTheta: \mathbb{P}^1(\mathbb{F}_{7^3}) \longrightarrow Grass( 3, 6, \mathbb{F}_7 )
\end{displaymath}
realizujemy teraz następująco: 
\begin{displaymath}
 \mathbb{F}_{7^3} \simeq V^{+} \ni x \longmapsto \operatorname{span}_{\mathbb{F}_{7^3}} \{ x + \xi \} \subset \mathbb{F}_7^6
\end{displaymath}
\begin{displaymath}
 \infty \longmapsto V^{+} \subset \mathbb{F}_7^6.
\end{displaymath}

Ogólnie, każda podprzestrzeń jednowymiarowa nad ~$\mathbb{F}_{7^3}$~ jest odwzorowywana na podprzestrzeń wymiaru trzy 
nad ~$\mathbb{F}_7$~ za pomocą operacji 
\begin{displaymath}
 \mathbb{F}_7^6 \ni u, ~~~\operatorname{span}_{\mathbb{F}_{7^3}} \{ u \} \mapsto \operatorname{span}_{\mathbb{F}_7} \{ f_1\cdot u, f_2\cdot u, f_3\cdot u \}.
\end{displaymath}
Poniżej przedstawiono niektóre z funkcji realizujące obliczanie kodów. Funkcja \textit{tuple()} generuje potrzebne krotki, natomiast wywoływana przez nią funkcja \textit{act()} wykonuje odpowiednie operacja macierzowe:

\begin{verbatim}
void    tuple( int n, int k, int d,
 int (*f)( int*, int, int[_n][_k] ), int *a, int b[_n][_k])
 {
  if( d > 0 ) {
               int     j;
               for( j=0; j<n; ++j ) {
                       a[d-1]= j;
                       tuple( n, k, d-1, f, a, b );
               }
       } else
               f( a, k, b );
}

int act( int u[_k]) 
{
 int v[_n], w[_n], r[_n][_k];
 static unsigned long	l= 1UL;
 mulV3( v, Vplus, u );
 addV( w, v, xi0 );
 mul3U( r, w );
 print3U( r );
}
\end{verbatim}

W wyniku tych działań otrzymujemy poszukiwane kody:
\begin{verbatim}
   1. [6 5 1 0 0 0
       1 0 0 1 1 0
       0 0 0 0 0 1]

   2. [5 2 4 5 4 0
       6 2 6 3 4 0
       2 6 5 5 0 4]

   3. [6 0 5 0 6 0
       2 3 2 2 3 6
       1 4 6 5 0 4]
  ...

 344. [2 5 6 4 3 1
       3 4 5 6 0 0
       2 1 4 4 6 3]
\end{verbatim}
Kolejnym krokiem będzie analiza jakościowa otrzymanego kodu.

\footnotesize
\renewcommand\refname{LITERATURA}

\bigskip

\end{document}